\documentclass[11pt]{article}

\usepackage[final]{acl}

\usepackage{times}
\usepackage{latexsym}

\usepackage[T1]{fontenc}

\usepackage[utf8]{inputenc}

\usepackage{microtype}

\usepackage{inconsolata}

\usepackage{graphicx}
\usepackage{mathrsfs}

\usepackage{CJKutf8}

\usepackage{booktabs}
\usepackage{array}

\usepackage{enumitem}
\usepackage{amsmath}
\usepackage{xltabular}
\usepackage{cuted}
\usepackage{subcaption}

\title{ReCAST: Restoration-aware Cascaded Stage-wise Training for Obfuscated SMS Risk Classification}

\author{Jieyun Huang, Yi Shen\thanks{Corresponding authors.}, Kaikai Zhao,  Jiangze Yan, Wenjing Zhang, \\
  {\bfseries Ping Chen, Ning Wang, Zhaoxiang Liu, Kai Wang, Shiguo Lian\footnotemark[1]} \\
  Unicom Data Intelligence, China Unicom \\
  Data Science \& Artificial Intelligence Research Institute, China Unicom \\
  \texttt{\{huangjy115, sheny73, zhaokk3, yanjz17, liansg\}}@chinaunicom.cn}
  
\begin{document}
\maketitle
\begin{abstract}

Fraudulent messages sent via Short Message Service (SMS) are increasingly obfuscated to evade cost-conscious classifiers in production systems. In Chinese SMS, attackers can exploit a wide range of carefully crafted obfuscation strategies to hide risk-bearing phrases while preserving human readability, making direct classification brittle under real-world latency and throughput constraints. We propose ReCAST, a Restoration-aware Cascaded Stage-wise Training framework for robust obfuscated Chinese SMS classification. ReCAST distills a large teacher model’s de-obfuscation ability into a smaller deployable student model by supervising obfuscated span detection, obfuscation type prediction, and text restoration, and then uses the restoration-aware student for downstream risk classification. Experiments on an internally constructed real-world Chinese SMS benchmark show that ReCAST substantially improves classification performance over directly trained baselines under obfuscation. The results suggest that restoration-aware distillation offers a practical path toward robust SMS risk classification with smaller deployable models under production-oriented constraints.

\end{abstract}

\section{Introduction}


Short message service (SMS) remains widely used, but is also abused for fraud, gambling, illegal promotion, and other risky activities. Industrial SMS risk detection systems must classify massive streams under strict latency, throughput, and cost constraints, making cost-conscious classifiers common for online inference. However, these classifiers are increasingly challenged by intentionally obfuscated SMS designed to evade detection.

\begin{figure}[t]
\centering
\includegraphics[width=\columnwidth]{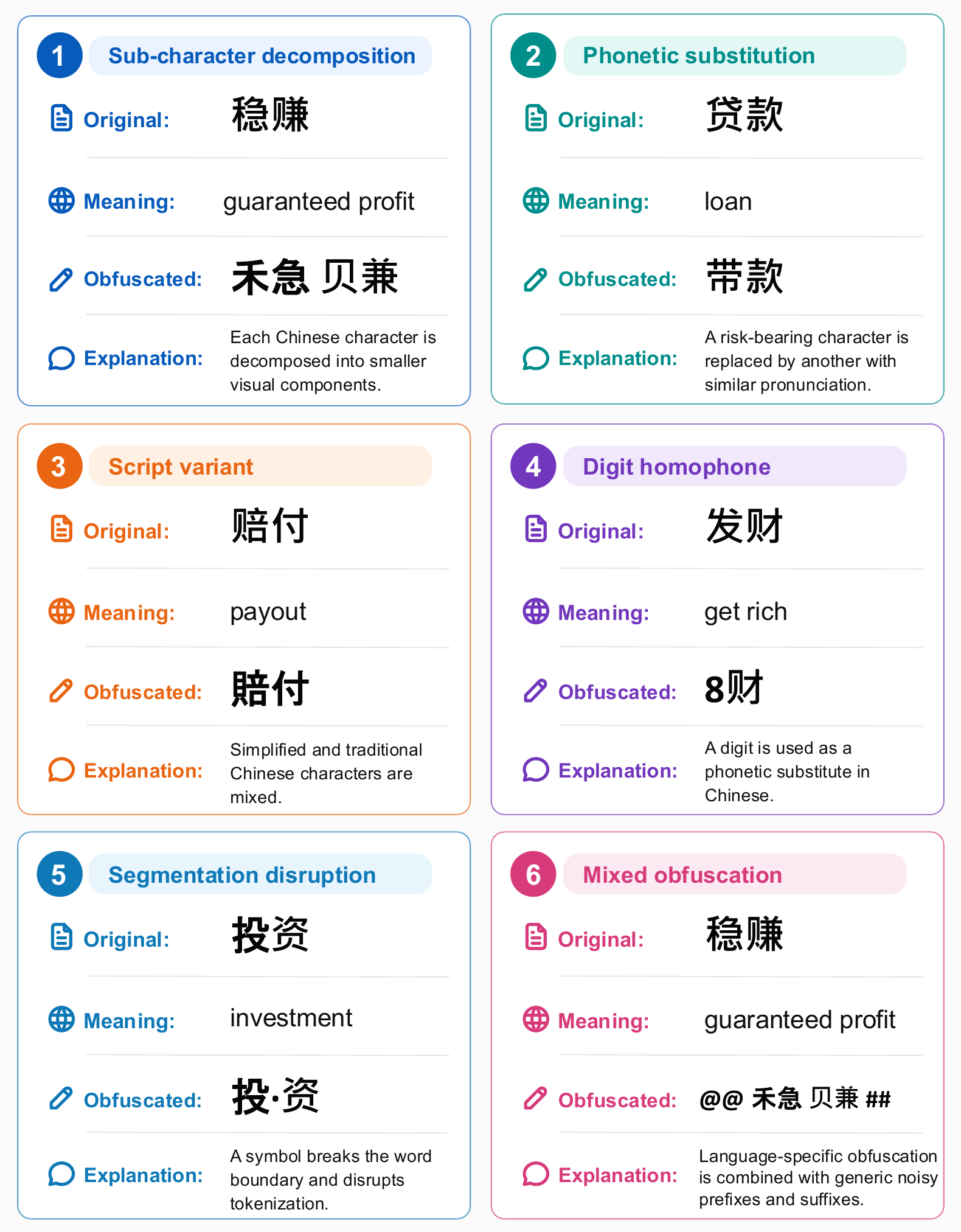}
\caption{Representative obfuscation strategies in Chinese SMS. Attackers hide risk-bearing phrases while preserving human readability with these strategies, making direct classification brittle. } \label{fig:obfuscate}
\end{figure}

Chinese SMS obfuscation is especially challenging because attackers can exploit the characteristics of the Chinese writing system and user reading habits. As shown in Figure 1, risk-bearing phrases can be hidden through carefully designed language-specific transformations, often combined with generic noisy perturbations, while remaining understandable to human recipients. \footnote{Please refer to Appendix \ref{app:obfuscation_list} for the complete Obfuscation Taxonomy.} These transformations preserve the intended meaning but remove, fragment, or distort the surface forms that classifiers, keyword rules, and tokenizers rely on, making the original risk indicators no longer directly observable in the input sequence.

This creates a practical robustness gap in deployed SMS risk detection systems. Rule-based methods and keyword matching are easily bypassed by surface-level modifications, while directly trained cost-conscious classifiers often fail when key risk indicators are transformed or fragmented. Large language models (LLMs) can often infer the intended meaning behind obfuscated text and restore normalized expressions, but invoking them for every incoming SMS is impractical in production due to latency, cost, and concurrency requirements. The central challenge is therefore how to transfer the de-obfuscation ability of LLMs into smaller models intended for online deployment.

To address this challenge, we propose ReCAST, a Restoration-aware Cascaded Stage-wise Training framework for robust classification of obfuscated Chinese SMS. ReCAST distills the de-obfuscation ability of a large teacher model into smaller deployable models without using the teacher during online inference. In the first stage, a student model learns structured restoration-oriented prediction, including obfuscated span detection, obfuscation type, and text restoration. In the second stage, the restoration-aware student is further trained for downstream risk classification. This design enables smaller deployable models to internalize de-obfuscation signals while maintaining single-pass inference efficiency in our serving setup.

We evaluate ReCAST on Chinese SMS datasets covering multiple risk categories. Compared with directly trained classification baselines, ReCAST achieves substantial improvements under obfuscation. Ablation studies confirm the effectiveness of restoration supervision and stage-wise training, while deployment-oriented evaluation shows that ReCAST preserves efficiency close to same-backbone online classification baselines. These results suggest that restoration-aware distillation provides a practical way to improve robustness against diverse real-world SMS obfuscation patterns.


Our contributions are summarized as follows:
\begin{itemize}[leftmargin=*, itemsep=1pt, topsep=2pt, parsep=0pt, partopsep=0pt]
    \item We characterize the industrial challenge of obfuscated Chinese SMS risk detection and discuss how various  obfuscation strategies can make classifiers brittle in real-world settings.
    \item We propose \textbf{ReCAST}, a restoration-aware Stage-wise distillation framework that transfers de-obfuscation ability from a large teacher model to cost-conscious classifiers.
    \item We demonstrate that ReCAST improves robustness under obfuscation while maintaining latency close to same-backbone single-pass classification baselines in a controlled serving setup.
\end{itemize}

\section{Related Work}

\textbf{SMS abuse detection under adaptive obfuscation.}
SMS abuse detection is commonly studied as a text-classification task, with recent approaches adopting BERT-style classifiers and LLM-based detectors \citep{liu2021spam, oswald2022spotspam, salman2025spallm}.
For industrial risk-control systems processing massive SMS streams under stringent latency, throughput, and cost constraints, lightweight LLM-based classifiers are more practical than directly serving large LLMs \citep{zhou2024survey}.
However, adaptive obfuscation challenges their robustness: malicious senders rewrite risk-bearing expressions into readable variants to evade detection \citep{hosseinpour2025evasive}.
This issue is particularly salient in Chinese SMS, where studies on Chinese spam detection show that homophonic, glyph-level, and sub-character variants can preserve readability while concealing risky content \citep{jiang2020signal, yao2022chinese, lai2022semorph}, and studies on Chinese offensive-language detection report similar cloaking effects \citep{xiao2024toxicloakcn, wu2025hedcold, guo2025lost}.

\textbf{Restoration and intermediate supervision.}
Under such obfuscation, training with only final category labels forces the classifier to implicitly locate evasive spans, infer canonical risk expressions, and associate the restored meaning with the risk label, making it difficult to maintain robustness as camouflage patterns evolve.
One natural defense is to normalize or restore the input before prediction \citep{bitton2022adversarial}.
However, this two-stage pipeline introduces an additional inference-time step, which may increase latency and propagate restoration errors.
To avoid this extra inference-time step, a classifier can instead internalize obfuscation-handling capability through richer training supervision, following the broader paradigms of knowledge distillation \citep{hinton2015distilling}, learning with privileged information \citep{vapnik2009privileged}, annotator rationales \citep{zaidan2007annotator}, and reasoning-step distillation \citep{hsieh2023distilling, feng2024teaching}.
ReCAST follows this intermediate supervision view, but replaces open-ended rationales with task-specific de-obfuscation supervision: obfuscated spans, obfuscation types, and restored text.
Because these signals are used only during training, deployment remains single-pass: the model directly maps raw SMS messages to risk labels while benefiting from the de-obfuscation capability learned during training.

\section{Methodology}

\subsection{Task Formulation}

Given an input Chinese SMS message $x$, the goal is to predict its risk label $y \in \mathcal{Y}$, where $\mathcal{Y}$ includes fraud, gambling, pornography, and benign. In real-world settings, $x$ may contain obfuscated expressions that hide risk-bearing phrases while preserving human readability.

Besides the final label $y$, ReCAST introduces restoration-oriented intermediate supervision during training. For an obfuscated message, we define a set of obfuscated spans $S = \{s_i\}_{i=1}^m$,  $T = \{t_i\}_{i=1}^m$, where \(t_i\) denotes the obfuscation type associated with span \(s_i\), and the restored text $r$. These annotations describe where the message is obfuscated, how it is obfuscated, and what the normalized message should be. They are used only during training; at inference time, the model directly predicts $y$ from the original input $x$.

\subsection{Overview of ReCAST}

\begin{figure*}[t]
\centering
\includegraphics[width=0.98\textwidth]{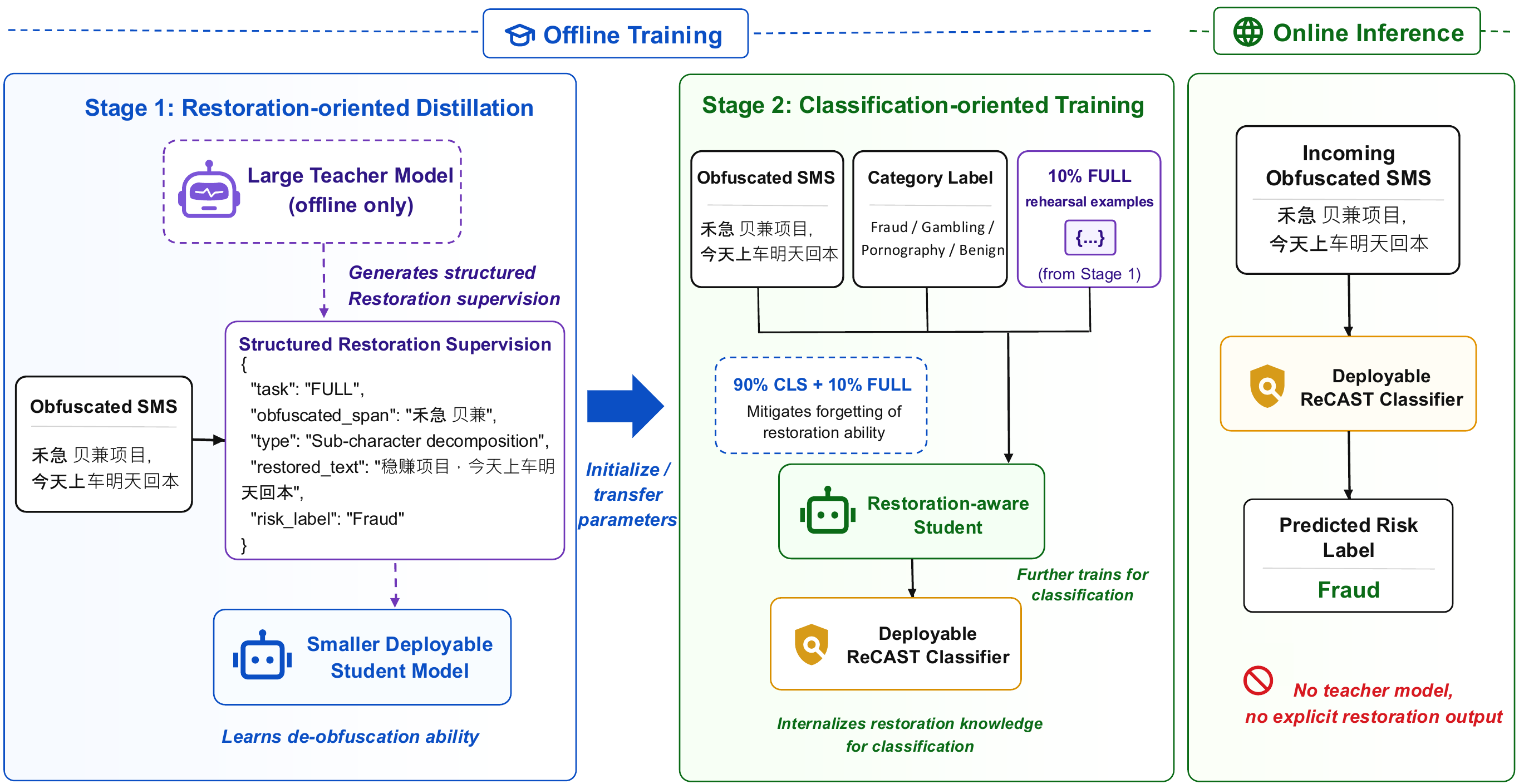}
\caption{ReCAST distills restoration-oriented supervision from a large teacher into a smaller deployable student, then further trains it for robust SMS risk classification. } \label{fig:RECAST}
\end{figure*}

We propose ReCAST, a Restoration-aware Cascaded Stage-wise Training framework for robust obfuscated SMS classification. ReCAST uses a large teacher model only during offline training, and distills its ability to interpret and restore obfuscated text into a smaller deployable student model for online deployment.

As shown in Figure~\ref{fig:RECAST}, ReCAST consists of two cascaded stages. Stage 1 trains the student with restoration-oriented supervision, including obfuscated spans, obfuscation types, restored text, and risk labels. Stage 2 further adapts the restoration-aware student to the downstream SMS risk classification task. During online inference, ReCAST takes the original SMS as input and directly outputs the risk label, without invoking the teacher model or generating the full restoration output.

A key design of ReCAST is a unified task-conditioned prompt interface. All tasks share the same system instruction and differ only in a task tag: \texttt{<TASK=FULL>} outputs structured restoration supervision and the risk label; \texttt{<TASK=CLS>} outputs only the final risk category; and \texttt{<TASK=NORM>} outputs only the restored text for explicit restoration baselines and diagnostic comparison. This unified interface makes Stage 1 and Stage 2 prompt-compatible, and allows Stage 2 to mix \texttt{FULL}-format rehearsal examples with \texttt{CLS}-format classification examples without changing the model interface. The complete prompt template is provided in Appendix~\ref{app:prompt}.

This design uses restoration as a training-time supervision signal rather than an inference-time preprocessing step, allowing the deployable classifier to benefit from de-obfuscation ability while preserving the efficiency required by production SMS risk detection systems.

\subsection{Stage 1: Restoration-oriented Distillation}

The first stage aims to teach the student model to interpret obfuscated SMS before optimizing it for final classification. Given an obfuscated SMS $x$, we use the large teacher model to produce restoration-oriented supervision. As shown in Figure~\ref{fig:RECAST}, this structured target is denoted as $(S,T,r,y)$, where $S$ represents obfuscated spans, $T$ represents obfuscation types, $r$ represents the restored SMS, and $y$ represents the risk category.

The student model is trained under the unified prompt interface with the \texttt{<TASK=FULL>} tag to generate the full structured output from the original obfuscated SMS. Unlike direct classification, this stage provides intermediate evidence that links obfuscated surface forms to their normalized risk-bearing expressions and corresponding obfuscation types. Such dense supervision encourages the student to learn mappings between obfuscated forms, restored expressions, and risk semantics.

In our implementation, structured labels are obtained from a combination of human annotations and LLM-assisted generation, followed by automatic checking and manual correction for flagged cases. The student is optimized with the standard sequence generation loss over the structured target output. After this stage, the resulting restoration-aware student has internalized preliminary de-obfuscation ability and is used to initialize the second-stage classifier. Complete examples of the structured output format are provided in Figure~\ref{fig:few-shot} in Appendix~\ref{app:prompt}.

\subsection{Stage 2: Classification-oriented Training}

In the second stage, we further train the restoration-aware student for SMS risk classification. The model initialized from Stage 1 has learned to associate obfuscated surface forms with their normalized expressions and risk semantics. Stage 2 transfers this knowledge from structured restoration generation to the final classification objective, so that the model can make robust label predictions without explicitly producing restoration outputs.

The input remains the original obfuscated SMS rather than the restored text, keeping the training setting consistent with online deployment, where ground-truth restorations are unavailable. We switch the task tag from \texttt{<TASK=FULL>} to \texttt{<TASK=CLS>} under the same unified prompt interface, and supervise the model to output only the final risk label $y$. This encourages the model to convert the explicit restoration behavior learned in Stage 1 into implicit decision evidence for classification, rather than relying on an external normalization step.

Because \texttt{CLS} and \texttt{FULL} examples share the same system prompt and differ only in the task tag and output schema, Stage 2 can naturally mix the two formats. Specifically, 90\% of the Stage 2 instances use the \texttt{<TASK=CLS>} format, while the remaining 10\% use the \texttt{<TASK=FULL>} format as rehearsal data. This lightweight replay strategy mitigates catastrophic forgetting of the restoration ability acquired in Stage 1, while allowing the training process to focus primarily on classification.

This design differs from an explicit restoration-then-classification pipeline. ReCAST does not first generate a restored SMS and then pass it to a separate classifier. Instead, restoration is used only as training-time supervision, and the resulting knowledge is internalized in the model parameters. After Stage 2, the model serves as the deployable ReCAST classifier and directly maps an incoming SMS to a risk category with the \texttt{<TASK=CLS>} tag.

\subsection{Inference}

During online inference, ReCAST uses only the deployable student classifier obtained after Stage 2. Given an incoming SMS, the model uses the same unified prompt interface with the \texttt{<TASK=CLS>} tag and directly predicts the risk label. It does not invoke the large teacher model, call a separate restoration module, or generate the full \texttt{FULL}-format output. This keeps inference as efficient as a single-pass cost-conscious classifier while still benefiting from the de-obfuscation knowledge distilled during training.

\section{Experiments}
\subsection{Experimental Setup}
\subsubsection{Datasets}

We use real-world Chinese SMS data for evaluation and a mixture of human-labeled and LLM-assisted data for training. 
The validation and test sets contain 500 and 1,000 production SMS messages, respectively, collected from different months to reduce temporal leakage. All evaluation samples are manually annotated with one of four labels: fraud, gambling, pornography, and benign. In these sets, non-benign messages include both obfuscated cases and a small number of unobfuscated cases. Benign messages do not contain intentional obfuscation, but may still include naturally occurring textual noise, such as typos, variant characters, or traditional Chinese forms, reflecting real-world SMS traffic. To prevent data leakage, validation and test messages are strictly excluded from all LLM-assisted training data generation steps, including seed selection, prompting, reference examples, and hard-example mining. All messages are anonymized: Personally identifiable information, phone numbers, account identifiers, and platform-specific signatures are removed or replaced with placeholders. The test set is approximately balanced across fraud, gambling, pornography, and benign categories, with each category accounting for about one quarter of the samples.

The training set contains 30K category-balanced examples constructed through a human-in-the-loop LLM-assisted pipeline. A subset is manually annotated with category labels and restored texts. For the rest, DeepSeek-V4-Pro\cite{deepseek2026v4} generates diverse clean SMS messages from seed examples and creates obfuscated variants using a small set of human-labeled obfuscation examples, covering both single and mixed strategies with at most two variants per clean message. We also mine hard examples by manually annotating a subset of LLM-misclassified obfuscated messages. Clean--obfuscated pairs are then converted into structured restoration-training instances with restored text, obfuscated spans, obfuscation types, and category labels. All generated instances are LLM-checked, with flagged samples manually reviewed and corrected. Details are provided in  Appendix \ref{app:dataset}.

\subsubsection{Baselines}

We compare ReCAST with four baselines covering direct classification, data augmentation, explicit restoration pipelines, and prompt-based inference. Unless otherwise specified, all trainable baselines use the same backbone as ReCAST.

\textbf{Direct-CLS} trains the backbone on the 30K training set using the original SMS as input and the risk label as output, without any restoration-oriented supervision.

\textbf{Aug-CLS} tests whether the gains come merely from more obfuscated data. It extends Direct-CLS with 20K additional synthetic obfuscated SMS messages, yielding 50K training examples, but still uses only classification labels.

\textbf{Pipeline-CLS} evaluates an explicit restoration-then-classification pipeline. We train a separate restoration model using the ReCAST training data, but only keep the original obfuscated SMS and its restored text as supervision. The model is trained with a \textbf{<TASK=NORM>} tag to normalize obfuscated SMS into restored text. The restored output is then fed into Direct-CLS for risk classification. This baseline contrasts with ReCAST, which uses restoration as training-time intermediate supervision and internalizes the restoration ability into a single classifier.

\textbf{Prompt-CLS} employs Qwen3.5-9B\cite{qwen3.5} without task-specific fine-tuning, using only a classification prompt. It serves as a zero-shot reference for the benefit of supervised distillation and fine-tuning.

\textbf{Published baselines.} To validate effectiveness beyond same-backbone controls, we additionally evaluate two published methods under the same SMS test protocol: \textbf{RoCBert} \citep{su-etal-2022-rocbert}, a robust Chinese encoder (RoCBert-base, $\approx$0.1B params) that incorporates semantic, phonetic, and visual information, fine-tuned for our four-way task; and an \textbf{adapted CA-CoT} \citep{yang-etal-2025-exploring-multimodal}, which explicitly identifies and restores linguistic perturbations before classification, using the same Qwen3.5-9B backbone as ReCAST while retaining CA-CoT's original three-stage reasoning procedure. We also report \textbf{DS-CLS}, direct inference of the DeepSeek-V4-Pro teacher with the classification prompt, as a non-deployable offline reference.

For Direct-CLS, Aug-CLS and Prompt-CLS, we use a direct classification prompt that asks the model to output only one risk category. The English translation of the prompt template is provided in Appendix \ref{app:prompt}, Figure \ref{fig:en-cls-prompt}.

\subsubsection{Evaluation Metrics}

We evaluate both classification effectiveness and serving efficiency.
For effectiveness, we report Accuracy (ACC) for overall four-way
classification and Risk Recall (RR) for recall over all non-benign categories,
i.e., the proportion of risky messages not predicted as benign. RR is
particularly important in production SMS risk detection, where missed risky
messages may cause downstream harm. For efficiency, we report Average Latency (Avg Lat.), P95 Latency (P95 Lat.), and Test Time (TT) under the same vLLM\cite{kwon2023efficient} serving configuration. Avg Lat. and P95 Lat. measure the mean and 95th-percentile end-to-end request latency, respectively, while TT measures the total wall-clock time for processing the 1,000-message test set. All efficiency metrics are averaged over multiple runs
with identical hardware, concurrency, batch-size, decoding, and prefix-caching settings.

\subsubsection{Implementation}
We use Qwen3.5-9B as the deployable student backbone for ReCAST and all same-backbone trainable baselines in Table 1, which is substantially smaller than the teacher model and avoids online teacher inference. All methods are trained under comparable configurations. Classification outputs are generated with constrained label decoding over the predefined label set. Detailed training hyperparameters and serving configurations are provided in Appendix \ref{app:implementation_details}.

\subsection{Main Results}


\begin{table}[t]
\centering
\small
\setlength{\tabcolsep}{2pt}
\begin{tabular*}{\columnwidth}{@{\extracolsep{\fill}}lccccc@{}}
\toprule
\textbf{Method}
& \textbf{ACC}
& \textbf{RR}
& \textbf{Avg}
& \textbf{P95}
& \textbf{TT}  \\
& (\%) & (\%) & (s) & (s) & (s) \\
\midrule
Prompt-CLS & 62.8 & 69.5 & 0.65 & 1.15 & 2.94 \\
Direct-CLS & 75.8\,$\pm$\,0.30 & 79.2\,$\pm$\,0.36 & 0.67 & 1.18 & 2.98 \\
Aug-CLS    & 80.6\,$\pm$\,0.28 & 85.1\,$\pm$\,0.31 & 0.68 & 1.17 & 3.00 \\
Pipeline-CLS & 83.4\,$\pm$\,0.53 & 86.2\,$\pm$\,0.47 & 1.42 & 2.52 & 6.66 \\

\textbf{ReCAST}
              & \textbf{86.6\,$\pm$\,0.17} & \textbf{89.5\,$\pm$\,0.15}
              & 0.71 & 1.20 & 3.03 \\
\bottomrule
\end{tabular*}
\caption{
Main results on internal real-world Chinese SMS test set. Trainable
methods report mean $\pm$ std over 4 random seeds; Prompt-CLS uses
deterministic constrained decoding and is a single value. Avg/P95/TT
are average latency, P95 latency, and total test time under 256
concurrent requests with the same vLLM serving configuration and
prefix caching.
}
\label{tab:main_results}
\end{table}

\begin{table}[t]
\centering
\small
\setlength{\tabcolsep}{3pt}
\begin{tabular}{llccc}
\toprule
\textbf{Method} & \textbf{Backbone} & \textbf{Params} & \textbf{ACC} & \textbf{RR} \\
& & & (\%) & (\%) \\
\midrule
RoCBert        & RoCBert-base   & $\approx$0.1B & 79.1 & 82.7 \\
Adapted CA-CoT & Qwen3.5-9B     & 9B            & 81.5 & 84.3 \\
\textbf{ReCAST}          & Qwen3.5-9B     & 9B            & \textbf{86.6} & \textbf{89.5} \\
\midrule
DS-CLS        & DeepSeek-V4-Pro & 1.6T         & 92.2 & 94.1 \\
\bottomrule
\end{tabular}
\caption{
Comparison with published methods and the teacher. RoCBert
\citep{su-etal-2022-rocbert} and CA-CoT \citep{yang-etal-2025-exploring-multimodal} are evaluated
under the same SMS test protocol; ReCAST outperforms both, including the
same-backbone CA-CoT. ReCAST values are the 4-seed mean from
Table~\ref{tab:main_results} (std omitted for compactness).
DS-CLS is direct teacher inference, a
non-deployable offline reference; ReCAST recovers about 94\% of its
accuracy.
}
\label{tab:published}
\end{table}

Table~\ref{tab:main_results} reports the main results on the real-world
obfuscated Chinese SMS test set. ReCAST achieves the best classification
performance, with 86.6\,$\pm$\,0.17\% Accuracy and 89.5\,$\pm$\,0.15\% Risk Recall. All gaps below are computed on the 4-seed mean values.
Compared with
Direct-CLS, it improves Accuracy and Risk Recall by 10.8 and 10.3 points,
respectively, showing that training only with final category labels is
insufficient for robust classification under intentional obfuscation. Aug-CLS improves over Direct-CLS by adding synthetic obfuscated examples,
but still lags behind ReCAST by 6.0 points in Accuracy and 4.4 points in
Risk Recall. This indicates that ReCAST's gains are not merely due to
exposure to more obfuscated data, but also come from restoration-aware
intermediate supervision. Against the strongest same-backbone baseline, Pipeline-CLS, ReCAST improves mean ACC and RR by 3.2 and 3.3 points.
Cross-seed variation is small for all trainable methods, the mean-based
ranking is unchanged, and ReCAST attains the lowest standard deviation
(0.17/0.15), confirming that these conclusions are robust to training randomness. Pipeline-CLS further confirms the value of
restoration, but its explicit restoration step leads to much higher serving
cost: 1.42s average latency, 2.52s P95 latency, and 6.66s total test time.

Beyond same-backbone controls, Table~\ref{tab:published} compares ReCAST
with two published methods---RoCBert \citep{su-etal-2022-rocbert} and an adapted
CA-CoT \citep{yang-etal-2025-exploring-multimodal} that explicitly restores perturbations
before classification on the Qwen3.5-9B backbone---and with DS-CLS,
direct teacher inference. ReCAST outperforms the published methods,
improving over the same-backbone CA-CoT by 5.1 ACC and 5.2 RR points,
which supports internalizing restoration-oriented supervision rather than
performing explicit restoration at inference time; it recovers about
94\% of the teacher's accuracy (86.6 vs.\ 92.2) while using a substantially smaller 9B student instead of the 1.6T-parameter teacher.

In contrast, ReCAST preserves the efficiency of single-pass classifiers.
Its average latency, P95 latency, and total test time are 0.71s, 1.20s,
and 3.03s, respectively, close to Direct-CLS and Aug-CLS under the same
vLLM serving configuration. These results show that ReCAST internalizes
restoration ability during training and achieves a favorable robustness-efficiency trade-off without explicit restoration at inference time. Under our target serving configuration, the observed latency is compatible with our offline deployment-oriented evaluation target, supporting the practical feasibility of ReCAST for subsequent shadow deployment and online validation.

Additional benign false positive analysis in Appendix~\ref{app:bfp} further shows that ReCAST ties Pipeline-CLS for the lowest Benign FPR among the student methods reported in Table 4, indicating that the improvement of ReCAST does not come from simply over-predicting risky categories. Class-wise results and the confusion matrix (Appendix~\ref{app:classwise}) show that Fraud is the most challenging category (F1 82.6\%), with residual errors dominated by risky-to-benign misclassification rather than confusion among the risky categories.

\subsection{Ablation Study}

\begin{table}[htbp]
\centering
\small
\begin{tabular}{lcc}
\toprule
\textbf{Variant} & \textbf{ACC$\uparrow$} & \textbf{RR $\uparrow$}  \\
\midrule
\textbf{ReCAST full} & \textbf{86.4} & \textbf{89.7} \\
Joint-Shuffled & 84.1 & 86.8 \\
w/o FULL replay in Stage 2 & 84.5 & 87.4 \\
w/o CLS label in Stage 1 & 85.3 & 87.8 \\
\midrule
w/o Stage 2  & 82.9 & 85.6  \\
- w/o span\&type in Stage 1 & 73.8 & 72.1  \\
- w/o restored text in Stage 1  & 78.6 & 80.4  \\

\bottomrule
\end{tabular}
\caption{Ablation of ReCAST. The upper block evaluates variants with Stage 2 classification training. The lower block removes Stage 2 and uses the CLS label generated in Stage 1 for classification, where each indented variant further removes one Stage 1 restoration signal. span\&type denotes joint supervision of obfuscated spans and their types. Ablations use a single representative checkpoint---the same one underlying the class-wise, restoration, and perturbation analyses---whose ACC/RR fall within 0.2pp of the 4-seed mean in Table~\ref{tab:main_results}. Because the smallest ablation drop (1.1pp ACC / 1.9pp RR) exceeds the largest cross-seed std in Table~\ref{tab:main_results} (0.53pp for ACC, 0.47pp for RR) by over 2$\times$, suggesting that the observed ablation effects are larger than the typical cross-seed variation.}
\label{tab:ablation}
\end{table}

Table~\ref{tab:ablation} summarizes the contribution of each ReCAST component. Removing FULL-format replay in Stage 2 reduces Accuracy from 86.4\% to 84.5\%, suggesting that replaying a small amount of restoration-format data helps preserve the de-obfuscation ability learned in Stage 1. Removing the CLS label from Stage 1 also hurts performance, showing that task-aware label supervision helps align restoration with downstream risk semantics.

To isolate the effect of stage-wise supervision ordering, we further compare ReCAST with Joint-Shuffled, which uses the same initialization, total FULL/CLS exposure, and computational budget but randomly mixes the two formats throughout training. Joint-Shuffled achieves 84.1\% ACC and 86.8\% RR, 2.3 and 2.9 points below ReCAST, respectively. This result indicates that the gain is not explained solely by the amount of restoration and classification supervision; organizing the transition from structured restoration-oriented learning to classification-oriented adaptation also contributes to the final performance.

Without Stage 2, Accuracy drops to 82.9\%, indicating that restoration-oriented training still needs classification adaptation. Further removing span\&type supervision causes the largest drop, to 73.8\% Accuracy and 72.1\% Risk Recall, highlighting the importance of localized obfuscation supervision. Removing restored text supervision substantially degrades performance, confirming the value of recovering normalized risk-bearing expressions.

\subsection{Restoration Analysis}

Because restoration-oriented supervision is central to Stage 1, we directly evaluate whether the intermediate restoration capability is learned and retained after classification-oriented adaptation. We randomly sample 200 obfuscated risky messages from the test set and manually annotate their obfuscated spans, restored forms, and obfuscation types. Using the <TASK=FULL> interface, the checkpoint immediately after Stage 1 obtains 92.0\% span-set exact-match accuracy and 3.5\% normalized character-level edit distance. The final ReCAST checkpoint achieves 89.0\% and 4.9\%, respectively. Although Stage 2 introduces a moderate degradation in explicit restoration quality, the final model retains most of the restoration capability while optimizing for direct classification, providing direct evidence that the restoration knowledge learned in Stage 1 is not fully forgotten.

\section{Conclusion}
We presented ReCAST, a restoration-aware cascaded stage-wise training framework for robust obfuscated Chinese SMS classification. ReCAST distills de-obfuscation ability from a large teacher into a smaller deployable model via structured supervision over obfuscated spans, restored text, and task-aware labels, followed by classification fine-tuning. Experiments on real-world obfuscated Chinese SMS show that ReCAST outperforms direct classification, data augmentation, and explicit restoration pipelines, while preserving single-pass inference efficiency comparable to same-backbone classifiers. Ablations further show that localized obfuscation and restored-text supervision are key to the gains. Although evaluated on Chinese SMS, ReCAST is conceptually applicable to other harmful text detection settings involving adversarial obfuscation, with broader validation left for future work.

\section*{Limitations}
This work has three main limitations. First, our experiments focus on Chinese SMS risk classification using a 9B-scale student backbone. Although ReCAST is not tied to a specific language, message channel, or model size, its effectiveness and efficiency on other languages, platforms, risk domains, and smaller deployable models remain to be systematically validated. Future work will explore whether restoration-aware supervision can provide similar robustness gains for more compact models under stricter latency and resource constraints.

Second, our training-data construction and structured supervision rely on a single teacher model, DeepSeek-V4-Pro. Although the test data are independently annotated by human experts and are excluded from all LLM-assisted training-data generation, the teacher may still introduce correlated biases in the types, styles, and coverage of synthetic obfuscation patterns. This may affect the representativeness of the resulting training corpus, particularly for naturally occurring, ambiguous, or evolving obfuscation patterns, so the robustness gains may not fully generalize, and the sensitivity of ReCAST to teacher choice remains an open question. As a sanity check, ReCAST achieves a benign false positive rate of 2.8\% on the held-out production test set; however, this does not establish fairness across all legitimate sender types or linguistic varieties.

Third, our deployment-oriented evaluation is conducted offline under controlled settings. It has not yet been evaluated under full production traffic with evolving templates, sender behaviors, and system-level constraints. More comprehensive shadow deployment or online monitoring is needed before high-impact operational use.

\section*{Ethics Statement}

This work aims to improve the robustness of SMS risk detection systems against obfuscated messages. Since SMS data may contain sensitive content, all data used in this study was processed under internal data governance procedures for security research and risk-control. Personally identifiable information, including phone numbers, account identifiers, verification codes, URLs, names, addresses, and sensitive fields, was removed or replaced with placeholders before training and evaluation. We do not release raw SMS messages or data that could be linked to individual users.

The teacher model used for training-data construction was deployed privately using the officially released DeepSeek-V4-Pro model weights rather than accessed through a third-party or public API. The released model weights and code are distributed under the MIT License. We use only the teacher's final structured outputs as offline supervision for training the student model and do not use intermediate reasoning traces as training targets.

We recognize that false positives may affect legitimate communication, such as platform notifications, verification messages, financial reminders, marketing messages, and ordinary personal messages. ReCAST has not been deployed as an autonomous production decision system. Rather, it is designed as an assistive risk-detection component. If deployed in practice, we recommend that its predictions be combined with rule-based safeguards, trusted sender signals, threshold control, manual review for high-impact cases, and appeal or correction mechanisms where appropriate.

This work also has potential dual-use risks because it analyzes obfuscation strategies used to evade SMS risk detectors. To mitigate misuse, we only present sanitized examples and avoid disclosing production rules, operational thresholds, complete evasion dictionaries, raw abusive messages, or deployment-sensitive prompts. The purpose of this work is defensive: improving anti-abuse systems against evolving real-world obfuscation.

For human annotation and review, the data were handled under existing internal security, risk-control, and data-governance procedures. Before research use and human annotation, personally identifiable information and other sensitive fields were removed or replaced. Annotation was conducted exclusively through an access-controlled internal system by authorized personnel bound by confidentiality agreements, who were provided with task guidelines and quality-control procedures and instructed to focus on risk categories and obfuscation patterns. Raw messages were not downloaded, exported, or transferred outside this controlled environment, within which DeepSeek-V4-Pro was also privately deployed. The restored text produced by our model should be treated as an intermediate analytical signal rather than factual evidence of user intent. If used in operational settings, the system would require periodic monitoring, evaluation, and updating, and should remain part of a broader human- and policy-governed anti-abuse pipeline.

\section*{Acknowledgments}

This work was supported by the National Natural Science Foundation of China Enterprise Innovation and Development Joint Fund Project U24B20179.

\bibliography{acl_latex}

@article{liu2021spam,
  title={A Spam Transformer Model for SMS Spam Detection},
  author={Liu, Xiaoxu and Lu, Haoye and Nayak, Amiya},
  journal={IEEE Access},
  volume={9},
  pages={80253--80263},
  year={2021},
  publisher={IEEE},
  doi={10.1109/ACCESS.2021.3081479},
  url={https://doi.org/10.1109/ACCESS.2021.3081479}
}

@article{salman2025spallm,
  title={SpaLLM-Guard: Pairing SMS Spam Detection Using Open-source and Commercial LLMs},
  author={Salman, Muhammad and Ikram, Muhammad and Basta, Nardine and Kaafar, Mohamed Ali},
  journal={arXiv preprint arXiv:2501.04985},
  year={2025},
  eprint={2501.04985},
  archivePrefix={arXiv},
  primaryClass={cs.CR},
  url={https://arxiv.org/abs/2501.04985}
}

@article{hosseinpour2025evasive,
  title={POSTER: A Multi-Signal Model for Detecting Evasive Smishing},
  author={Hosseinpour, Shaghayegh and Das, Sanchari},
  journal={arXiv preprint arXiv:2505.18233},
  year={2025},
  eprint={2505.18233},
  archivePrefix={arXiv},
  primaryClass={cs.LG},
  url={https://arxiv.org/abs/2505.18233}
}

@inproceedings{xiao2024toxicloakcn,
  title={ToxiCloakCN: Evaluating Robustness of Offensive Language Detection in Chinese with Cloaking Perturbations},
  author={Xiao, Yunze and Hu, Yujia and Choo, Kenny Tsu Wei and Lee, Roy Ka-Wei},
  booktitle={Proceedings of the 2024 Conference on Empirical Methods in Natural Language Processing},
  pages={6012--6025},
  year={2024},
  address={Miami, Florida, USA},
  publisher={Association for Computational Linguistics},
  doi={10.18653/v1/2024.emnlp-main.345},
  url={https://aclanthology.org/2024.emnlp-main.345/}
}

@inproceedings{guo2025lost,
  title={Lost in Pronunciation: Detecting Chinese Offensive Language Disguised by Phonetic Cloaking Replacement},
  author={Guo, Haotan and He, Jianfei and Ma, Jiayuan and Na, Hongbin and Wang, Zimu and Zhang, Haiyang and Chen, Qi and Wang, Wei and Shi, Zijing and Shen, Tao and Chen, Ling},
  booktitle={Proceedings of the 2025 Conference on Empirical Methods in Natural Language Processing: Industry Track},
  pages={2538--2550},
  year={2025},
  address={Suzhou, China},
  publisher={Association for Computational Linguistics},
  doi={10.18653/v1/2025.emnlp-industry.172},
  url={https://aclanthology.org/2025.emnlp-industry.172/}
}

@inproceedings{jiang2020signal,
  title={Camouflaged Chinese Spam Content Detection with Semi-supervised Generative Active Learning},
  author={Jiang, Zhuoren and Gao, Zhe and Duan, Yu and Kang, Yangyang and Sun, Changlong and Zhang, Qiong and Liu, Xiaozhong},
  booktitle={Proceedings of the 58th Annual Meeting of the Association for Computational Linguistics},
  pages={3080--3085},
  year={2020},
  address={Online},
  publisher={Association for Computational Linguistics},
  doi={10.18653/v1/2020.acl-main.279},
  url={https://aclanthology.org/2020.acl-main.279/}
}

@article{yao2022chinese,
  title={Chinese Spam Detection Using a Hybrid BiGRU-CNN Network with Joint Textual and Phonetic Embedding},
  author={Yao, Jinliang and Wang, Chenrui and Hu, Chuang and Huang, Xiaoxi},
  journal={Electronics},
  volume={11},
  number={15},
  pages={2418},
  year={2022},
  publisher={MDPI},
  doi={10.3390/electronics11152418},
  url={https://www.mdpi.com/2079-9292/11/15/2418}
}

@inproceedings{wu2025hedcold,
  title={Enhancing Chinese Offensive Language Detection with Homophonic Perturbation},
  author={Wu, Junqi and Ji, Shujie and Zhong, Kang and Peng, Huiling and {Zhendongxiao} and Liu, Xiongding and Wei, Wu},
  booktitle={Proceedings of the 2025 Conference on Empirical Methods in Natural Language Processing},
  pages={22660--22675},
  year={2025},
  address={Suzhou, China},
  publisher={Association for Computational Linguistics},
  doi={10.18653/v1/2025.emnlp-main.1154},
  url={https://aclanthology.org/2025.emnlp-main.1154/}
}

@inproceedings{bitton2022adversarial,
  title={Adversarial Text Normalization},
  author={Bitton, Joanna and Pavlova, Maya and Evtimov, Ivan},
  booktitle={Proceedings of the 2022 Conference of the North American Chapter of the Association for Computational Linguistics: Human Language Technologies: Industry Track},
  pages={268--279},
  year={2022},
  address={Hybrid: Seattle, Washington + Online},
  publisher={Association for Computational Linguistics},
  doi={10.18653/v1/2022.naacl-industry.30},
  url={https://aclanthology.org/2022.naacl-industry.30/}
}

@inproceedings{feng2024teaching,
    title = "Teaching Small Language Models Reasoning through Counterfactual Distillation",
    author = "Feng, Tao  and
      Li, Yicheng  and
      Li, Chenglin  and
      Chen, Hao  and
      Yu, Fei  and
      Zhang, Yin",
    editor = "Al-Onaizan, Yaser  and
      Bansal, Mohit  and
      Chen, Yun-Nung",
    booktitle = "Proceedings of the 2024 Conference on Empirical Methods in Natural Language Processing",
    month = nov,
    year = "2024",
    address = "Miami, Florida, USA",
    publisher = "Association for Computational Linguistics",
    url = "https://aclanthology.org/2024.emnlp-main.333/",
    doi = "10.18653/v1/2024.emnlp-main.333",
    pages = "5831--5842"
}

@inproceedings{hsieh2023distilling,
    title = "Distilling Step-by-Step! Outperforming Larger Language Models with Less Training Data and Smaller Model Sizes",
    author = "Hsieh, Cheng-Yu  and
      Li, Chun-Liang  and
      Yeh, Chih-kuan  and
      Nakhost, Hootan  and
      Fujii, Yasuhisa  and
      Ratner, Alex  and
      Krishna, Ranjay  and
      Lee, Chen-Yu  and
      Pfister, Tomas",
    editor = "Rogers, Anna  and
      Boyd-Graber, Jordan  and
      Okazaki, Naoaki",
    booktitle = "Findings of the Association for Computational Linguistics: ACL 2023",
    month = jul,
    year = "2023",
    address = "Toronto, Canada",
    publisher = "Association for Computational Linguistics",
    url = "https://aclanthology.org/2023.findings-acl.507/",
    doi = "10.18653/v1/2023.findings-acl.507",
    pages = "8003--8017"
}

@article{oswald2022spotspam,
  title={SpotSpam: Intention Analysis--driven SMS Spam Detection Using BERT Embeddings},
  author={Oswald, C. and Simon, Sona Elza and Bhattacharya, Arnab},
  journal={ACM Transactions on the Web},
  volume={16},
  number={3},
  pages={1--27},
  year={2022},
  publisher={Association for Computing Machinery},
  doi={10.1145/3538491}
}

@inproceedings{lai2022semorph,
  title={Semorph: A Morphology Semantic Enhanced Pre-trained Model for Chinese Spam Text Detection},
  author={Lai, Kaiting and Long, Yinong and Wu, Bowen and Li, Ying and Wang, Baoxun},
  booktitle={Proceedings of the 31st ACM International Conference on Information \& Knowledge Management},
  pages={1003--1013},
  year={2022},
  publisher={ACM},
  doi={10.1145/3511808.3557448}
}

@misc{hinton2015distilling,
  title={Distilling the Knowledge in a Neural Network},
  author={Hinton, Geoffrey and Vinyals, Oriol and Dean, Jeff},
  year={2015},
  eprint={1503.02531},
  archivePrefix={arXiv},
  primaryClass={stat.ML},
  url={https://arxiv.org/abs/1503.02531}
}

@article{vapnik2009privileged,
  title={A New Learning Paradigm: Learning Using Privileged Information},
  author={Vapnik, Vladimir and Vashist, Akshay},
  journal={Neural Networks},
  volume={22},
  number={5--6},
  pages={544--557},
  year={2009},
  publisher={Elsevier},
  doi={10.1016/j.neunet.2009.06.042}
}

@inproceedings{zaidan2007annotator,
  title={Using ``Annotator Rationales'' to Improve Machine Learning for Text Categorization},
  author={Zaidan, Omar and Eisner, Jason and Piatko, Christine},
  booktitle={Human Language Technologies 2007: The Conference of the North American Chapter of the Association for Computational Linguistics; Proceedings of the Main Conference},
  pages={260--267},
  year={2007},
  publisher={Association for Computational Linguistics},
  url={https://aclanthology.org/N07-1033/}
}

@misc{zhou2024survey,
  title={A Survey on Efficient Inference for {Large Language Models}},
  author={Zhou, Zixuan and Ning, Xuefei and Hong, Ke and Fu, Tianyu and Xu, Jiaming and Li, Shiyao and Lou, Yuming and Wang, Luning and Yuan, Zhihang and Li, Xiuhong and Yan, Shengen and Dai, Guohao and Zhang, Xiao-Ping and Dong, Yuhan and Wang, Yu},
  year={2024},
  eprint={2404.14294},
  archivePrefix={arXiv},
  primaryClass={cs.CL},
  url={https://arxiv.org/abs/2404.14294}
}

@inproceedings{su-etal-2022-rocbert,
    title = "{R}o{CB}ert: Robust {C}hinese Bert with Multimodal Contrastive Pretraining",
    author = "Su, Hui  and
      Shi, Weiwei  and
      Shen, Xiaoyu  and
      Xiao, Zhou  and
      Ji, Tuo  and
      Fang, Jiarui  and
      Zhou, Jie",
    editor = "Muresan, Smaranda  and
      Nakov, Preslav  and
      Villavicencio, Aline",
    booktitle = "Proceedings of the 60th Annual Meeting of the Association for Computational Linguistics (Volume 1: Long Papers)",
    month = may,
    year = "2022",
    address = "Dublin, Ireland",
    publisher = "Association for Computational Linguistics",
    url = "https://aclanthology.org/2022.acl-long.65/",
    doi = "10.18653/v1/2022.acl-long.65",
    pages = "921--931"
}

@inproceedings{yang-etal-2025-exploring-multimodal,
    title = "Exploring Multimodal Challenges in Toxic {C}hinese Detection: Taxonomy, Benchmark, and Findings",
    author = "Yang, Shujian  and
      Cui, Shiyao  and
      Hu, Chuanrui  and
      Wang, Haicheng  and
      Zhang, Tianwei  and
      Huang, Minlie  and
      Lu, Jialiang  and
      Qiu, Han",
    editor = "Che, Wanxiang  and
      Nabende, Joyce  and
      Shutova, Ekaterina  and
      Pilehvar, Mohammad Taher",
    booktitle = "Findings of the Association for Computational Linguistics: ACL 2025",
    month = jul,
    year = "2025",
    address = "Vienna, Austria",
    publisher = "Association for Computational Linguistics",
    url = "https://aclanthology.org/2025.findings-acl.742/",
    doi = "10.18653/v1/2025.findings-acl.742",
    pages = "14382--14396",
    ISBN = "979-8-89176-256-5"
}

@inproceedings{kwon2023efficient,
author = {Kwon, Woosuk and Li, Zhuohan and Zhuang, Siyuan and Sheng, Ying and Zheng, Lianmin and Yu, Cody Hao and Gonzalez, Joseph and Zhang, Hao and Stoica, Ion},
title = {Efficient Memory Management for Large Language Model Serving with PagedAttention},
year = {2023},
isbn = {9798400702297},
publisher = {Association for Computing Machinery},
address = {New York, NY, USA},
url = {https://doi.org/10.1145/3600006.3613165},
doi = {10.1145/3600006.3613165},
booktitle = {Proceedings of the 29th Symposium on Operating Systems Principles},
pages = {611–626},
numpages = {16},
location = {Koblenz, Germany},
series = {SOSP '23}
}

@techreport{deepseek2026v4,
  title = {Deepseek-v4: Towards highly efficient million-token context intelligence},
  author = {{DeepSeek-AI}},
  year = {2026},
  url = {https://huggingface.co/deepseek-ai/DeepSeek-V4-Pro/blob/main/DeepSeek_V4.pdf}
}

@misc{qwen3.5,
  title = {{Qwen3.5}: Towards Native Multimodal Agents},
  author = {{Qwen Team}},
  month = {February},
  year = {2026},
  url = {https://qwen.ai/blog?id=qwen3.5}
}

\clearpage

\appendix

\section{Dataset Details}
\label{app:dataset}

\subsection{Evaluation Data}

The validation and test sets are sampled from real production SMS data and manually annotated by human experts. The validation set contains 500 messages, while the test set contains 1,000 messages. The two sets are collected from different months to reduce temporal leakage and to better evaluate robustness under temporal distribution shift. Each message is assigned one of four category labels: fraud, gambling, pornography and benign. The evaluation sets are never generated or rewritten by the LLM.

\subsection{Training Data Construction}
The final training set contains 30K examples with approximately balanced category distributions. It is built from both human-labeled data and LLM-assisted synthetic data. A subset of the training data is manually annotated with risk categories and restored texts. To scale up the training data, we use DeepSeek-V4-Pro in a multi-step generation pipeline.

First, we provide a set of seed examples to the LLM and ask it to generate clean SMS messages with similar topics. We explicitly require the generated messages to be category-balanced, diverse in topic, sentence pattern, and length, and free from near duplicates. These clean messages serve as the base texts for later obfuscation.

Second, we provide a small number of human-labeled obfuscation examples and ask the LLM to create obfuscated variants from the clean messages. The generated variants cover all single obfuscation strategies in our taxonomy, including homophone substitution, visually similar characters, simplified--traditional variants, digit homophones, whitespace insertion, symbol insertion, URL splitting, prefix/suffix noise, abbreviation, slang, and sub-character decomposition. We also generate mixed obfuscations by combining multiple strategies. To avoid over-representing a single clean message, each clean message is used to generate at most two obfuscated variants.

Third, we perform LLM-assisted hard example mining. We ask DeepSeek-V4-Pro to directly classify a subset of obfuscated messages and select a subset of the incorrectly predicted samples for manual annotation. Human annotators then provide corrected labels and restored texts for these hard cases. This step enriches the training set with examples that are likely to confuse direct classification models.

Fourth, we prompt the LLM to convert clean--obfuscated pairs into structured restoration-training instances. Each instance contains the obfuscated SMS, restored text, risk category, obfuscated spans, and corresponding obfuscation types in JSON format. These structured labels are used to supervise the restoration stage of ReCAST.

\subsection{Quality Control}
We apply both automatic and manual quality control to the generated training data. The LLM is first used to check whether each generated instance is internally consistent, including whether the restored text matches the obfuscated text, whether the labeled span corresponds to an actual obfuscated fragment, whether the obfuscation type is appropriate, and whether the category label is plausible. Instances flagged as potentially problematic are manually reviewed and corrected before being added to the final training set.

We also control data diversity during generation by requiring diverse sentence patterns, lengths, and topics, and by limiting the number of obfuscated variants generated from each clean message. This reduces duplication and prevents the model from overfitting to a small set of templates. The resulting training set provides large-scale structured supervision for restoration-aware distillation, while all validation and test results are reported on real, manually annotated production SMS data.

\section{Implementation Details}
\label{app:implementation_details}
\textbf{Model and Hardware.} We adopt Qwen3.5-9B as our backbone model. Full-parameter fine-tuning is conducted using the \textsc{Swift} framework on eight NVIDIA H100 (80GB) GPUs in \text{BF16} precision. 

\textbf{Hyperparameters.} Across all stages, we employ the AdamW optimizer with a learning rate of $1 \times 10^{-5}$, a weight decay of 0.1, and a linear learning rate scheduler with a 0.05 warmup ratio. The training process spans 5 epochs with a maximum sequence length of 4096 tokens. With a per-device batch size of 2 and 16 gradient accumulation steps, the effective batch size scales to 256.

\textbf{Inference Serving Setup.} 
We deploy the model using the vLLM serving engine on a single NVIDIA H100 GPU (tensor parallelism degree of 1). Serving parameters are configured with a GPU memory utilization of 0.93 and a maximum of 256 concurrent sequences, with prefix caching enabled to enhance serving throughput. 
The maximum generation length is capped at 256 tokens, sufficient for the outputs observed in our experiments. For classification evaluation, we use constrained label decoding, where the output is restricted to the predefined risk-label set, ensuring deterministic and valid category predictions.

\textbf{Teacher model and generation settings.}
We use a privately deployed DeepSeek-V4-Pro, a Mixture-of-Experts (MoE) model with 1.6T total parameters, as the offline teacher for synthetic data generation, hard-example mining, and restoration-oriented annotation. The teacher is substantially larger than the 9B student backbone and is used only during training-data construction. During teacher-side generation, we enable reasoning mode to improve annotation consistency for difficult obfuscation cases. We retain only the final structured outputs, including restored text, obfuscated spans, obfuscation types, and risk labels; intermediate reasoning traces are discarded and are not used as student supervision. We additionally evaluate direct teacher inference as an offline reference, but do not consider it a deployable baseline due to substantially higher serving cost. The results are shown in Table \ref{tab:BFP}.

\section{Benign False Positive Analysis}
\label{app:bfp}

In production SMS risk detection, improving recall on risky messages should not come at the cost of excessive false positives on benign traffic. We therefore report the benign false positive rate (Benign FPR), defined as the proportion of benign messages incorrectly predicted as fraud, gambling, or pornography. This metric complements Risk Recall by measuring whether a model tends to over-predict risky categories on normal SMS messages.


\begin{table}[t]
\centering
\small
\begin{tabular}{lcccl}
\toprule
\textbf{Method} & \textbf{ACC$\uparrow$} & \textbf{RR$\uparrow$} & \textbf{Benign FPR$\downarrow$}  \\

\midrule
DS-CLS & 92.2 & 94.1 & 2.4 \\

\midrule
Prompt-CLS & 62.8 & 69.5 & 16.0  \\
Direct-CLS & 75.7 & 79.6 & 6.8  \\
Aug-CLS & 80.3 & 85.5 & 6.4 \\
Pipeline-CLS & 82.9 & 86.2 & 2.8  \\
ReCAST & 86.4 & 89.7 & 2.8  \\
\bottomrule
\end{tabular}
\caption{Benign false positive analysis on the real-world Chinese SMS test set (representative ReCAST checkpoint). The benign subset contains 250 messages. Benign FPR measures the proportion of benign messages incorrectly classified as fraud, gambling, or pornography. DS-CLS directly applies DeepSeek-V4-Pro with the direct classification prompt and is included only as a non-deployable offline teacher reference.}
\label{tab:BFP}
\end{table}

As shown in Table~\ref{tab:BFP}, ReCAST achieves the best overall classification performance while maintaining the lowest benign false positive rate. Compared with Prompt-CLS, ReCAST substantially reduces Benign FPR from 16.0\% to 2.8\%, indicating that supervised restoration-aware training greatly improves robustness to benign but risk-looking surface patterns. ReCAST also achieves a lower Benign FPR than Direct-CLS and Aug-CLS, and matches Pipeline-CLS, while substantially outperforming Pipeline-CLS in ACC and Risk Recall. This suggests that ReCAST does not improve Risk Recall by simply over-predicting risky categories. Instead, restoration-aware supervision helps the model better distinguish intentional obfuscation in risky messages from naturally occurring noise in benign SMS.

We further manually inspect the benign false positives made by ReCAST. Most of the remaining benign errors are normal notifications related to banking, financial, or securities services. These messages often contain URLs, account-operation terms, amounts, or stock-code-like strings, which can resemble risk-bearing cues in fraudulent SMS. This indicates that the residual false positives are concentrated in high-risk-looking but legitimate transactional scenarios. In practical deployment, trusted-sender signals, domain or sender whitelists, and lightweight business rules can be combined with ReCAST to further reduce such false positives.

\section{Class-wise Analysis}
\label{app:classwise}

Table~\ref{tab:confusion} reports the confusion matrix of the ReCAST checkpoint on the 1,000-message test set, and Table~\ref{tab:classwise} reports the corresponding per-class precision, recall, and F1.

\begin{table}[t]
\centering
\small
\setlength{\tabcolsep}{4pt}
\begin{tabular}{lcccc}
\toprule
\textbf{Actual \textbackslash{} Pred.} & \textbf{Benign} & \textbf{Gambling} & \textbf{Porn.} & \textbf{Fraud} \\
\midrule
Benign      & 243 & 2  & 1 & 4  \\
Gambling    & 22  & 201 & 5 & 15 \\
Pornography & 17  & 5  & 220 & 7  \\
Fraud       & 38  & 12 & 8 & 200 \\
\bottomrule
\end{tabular}
\caption{Confusion matrix of the ReCAST checkpoint on the 1,000-message test set (rows: actual, columns: predicted).}
\label{tab:confusion}
\end{table}

\begin{table}[t]
\centering
\small
\begin{tabular}{lccc}
\toprule
\textbf{Category} & \textbf{Precision (\%)} & \textbf{Recall (\%)} & \textbf{F1 (\%)} \\
\midrule
Benign      & 75.9 & 97.2 & 85.3 \\
Gambling    & 91.4 & 82.7 & 86.8 \\
Pornography & 94.0 & 88.4 & 91.1 \\
Fraud       & 88.5 & 77.5 & 82.6 \\
\bottomrule
\end{tabular}
\caption{Class-wise precision, recall, and F1 of the ReCAST checkpoint. Pornography obtains the highest F1, while Fraud is the most challenging category.}
\label{tab:classwise}
\end{table}

Pornography obtains the highest F1 (91.1\%), whereas Fraud is the most challenging category (F1 82.6\%): 38 of 258 fraud messages are predicted as Benign, so the dominant residual error is risky-to-benign misclassification rather than confusion among the three risky categories. For Benign messages, ReCAST achieves 97.2\% recall, with only seven benign messages incorrectly predicted as risky; manual inspection shows that most of these false positives are legitimate banking, financial, or securities notifications whose URLs, account-operation terms, monetary amounts, or stock-code-like strings resemble risk-bearing cues in fraudulent SMS.


\onecolumn
\section{Unified Multi-task Prompt Template}
\label{app:prompt}


\begin{minipage}{\textwidth} 
\centering 
\includegraphics[width=1\textwidth, height=1.43\textwidth]{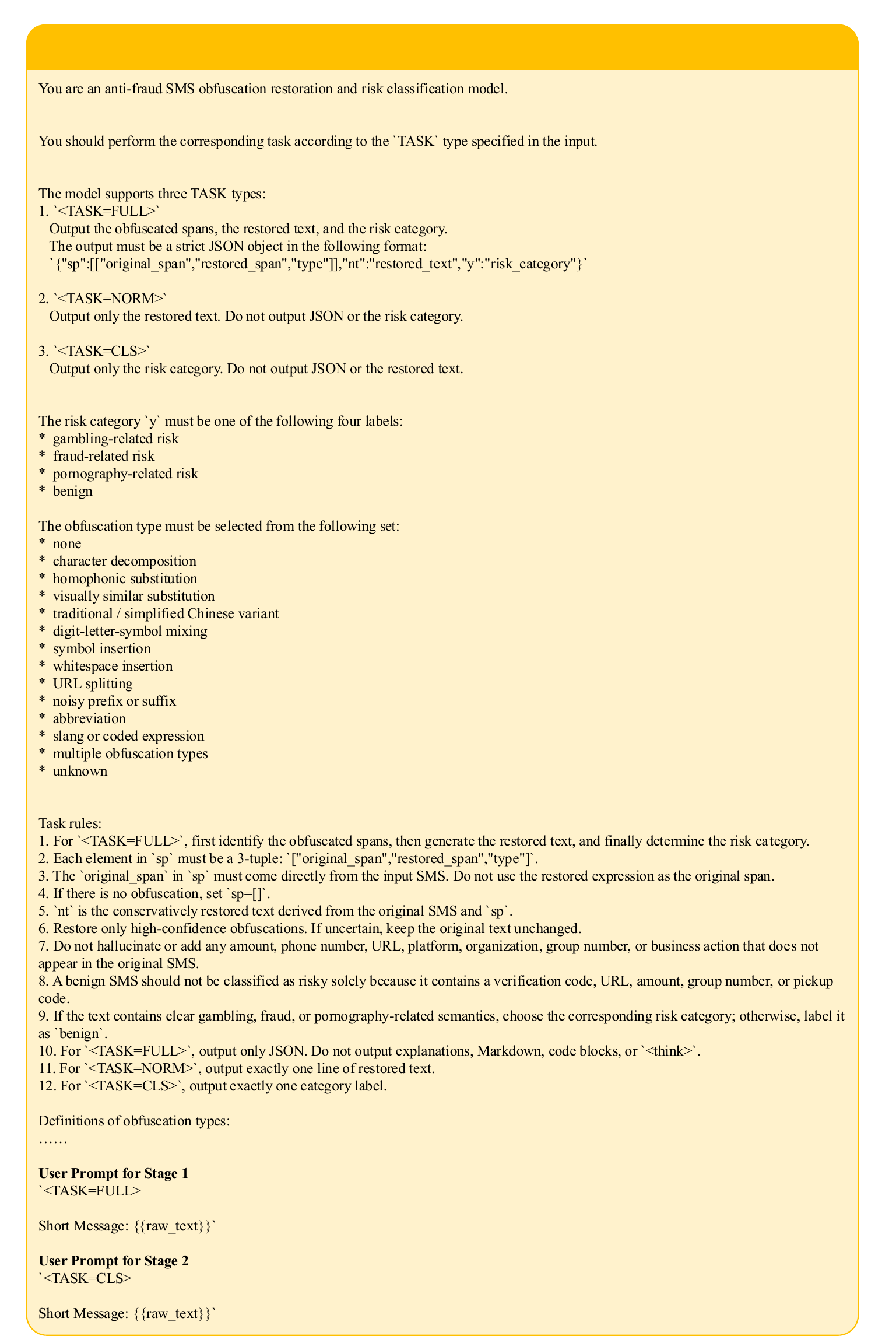} 
\captionof{figure}{English translation of the unified multi-task prompt template. Provided solely for readability; the original Chinese prompts are used for training and evaluation.}
\label{fig:prompt-en}
\end{minipage}

\begin{figure*}[!t]
\centering
\includegraphics[width=\textwidth]{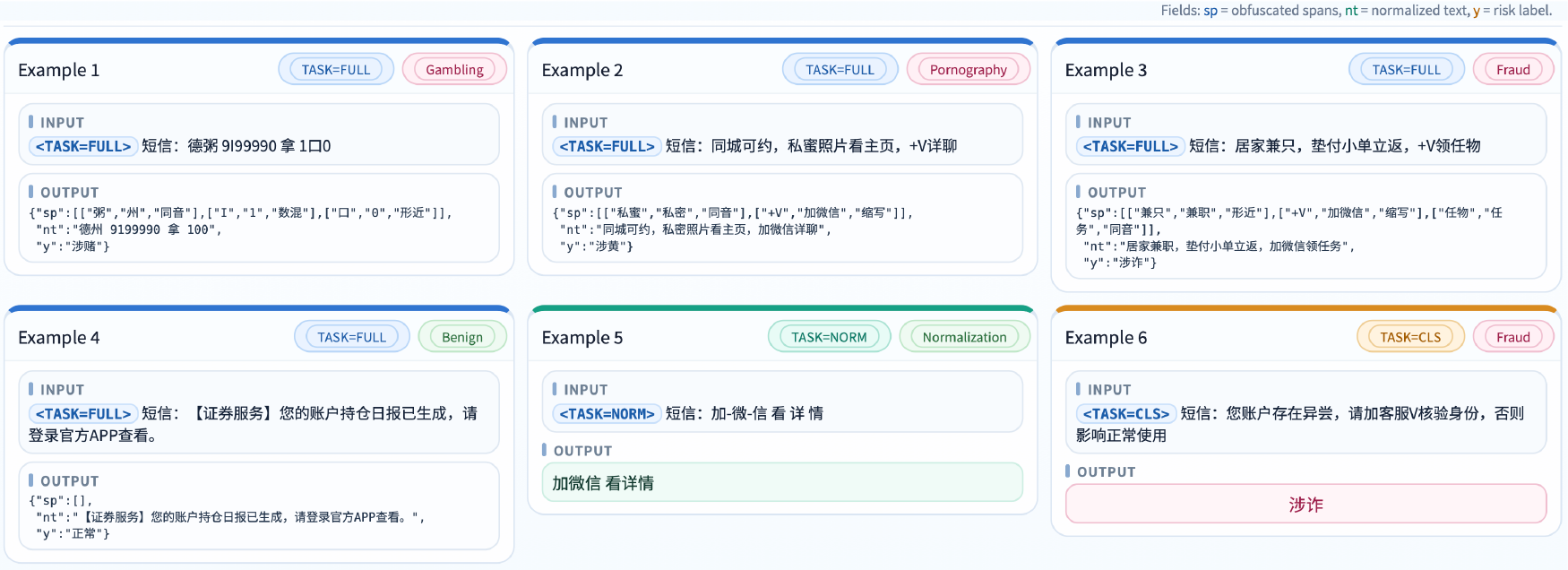}
\captionof{figure}{Few-Shot examples.} \label{fig:few-shot}
\end{figure*}


\begin{figure*}[!t]
\centering
\includegraphics[width=0.95\linewidth]{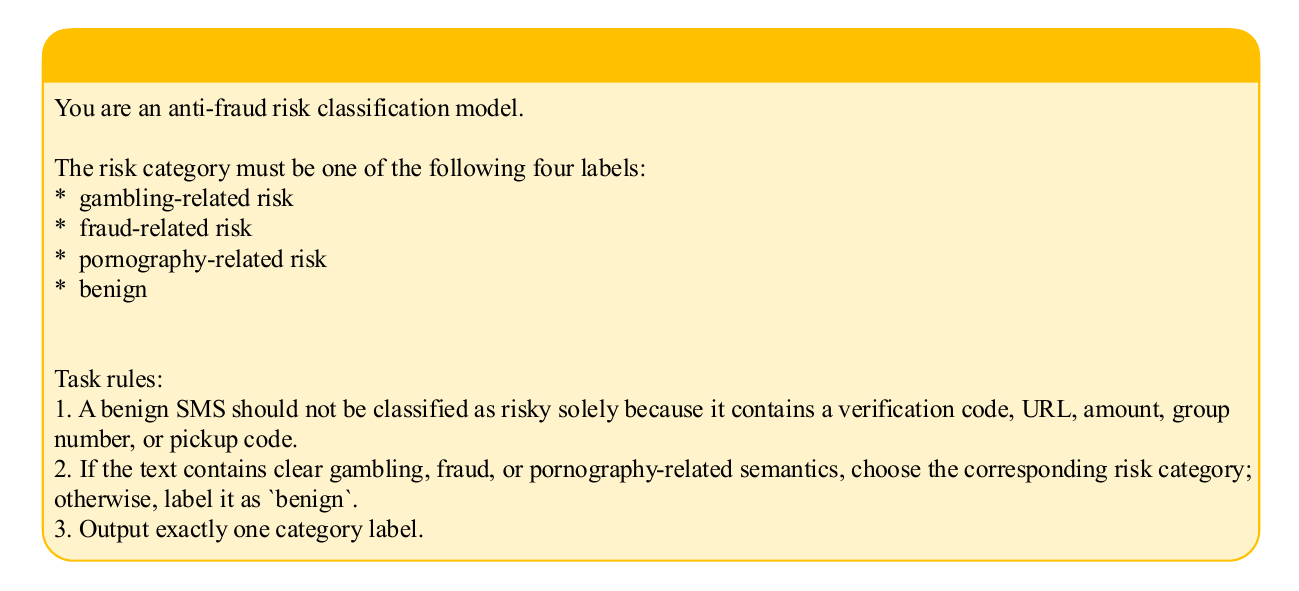}
\caption{This is an English translation of the direct classification prompt template for the Prompt‑CLS, Direct‑CLS, and Aug‑CLS baselines; it is provided for readability only, while the original Chinese prompt is used in all experiments.} 
\label{fig:en-cls-prompt}
\end{figure*}

\onecolumn 
\section{Obfuscation Type List}
\label{app:obfuscation_list}

\begin{CJK*}{UTF8}{gbsn}
\begin{xltabular}{\textwidth}{>{\hsize=0.3\hsize}X >{\hsize=0.9\hsize}X >
{\hsize=1.2\hsize}X >
{\hsize=1.4\hsize}X >{\hsize=1.2\hsize}X}

\caption{Taxonomy of obfuscation types used for structured de-obfuscation supervision.}
\label{tab:obfuscation_taxonomy_appendix} \\
\toprule
\textbf{Type} & \textbf{English Name} & \textbf{Chinese Explanation} & \textbf{English Explanation} & \textbf{Examples} \\
\midrule
\endfirsthead

\multicolumn{5}{c}{{\bfseries Table \thetable\ -- Continued from previous page}} \\
\toprule
\textbf{Type} & \textbf{English Name} & \textbf{Chinese Explanation} & \textbf{English Explanation} & \textbf{Examples} \\
\midrule
\endhead

\midrule
\multicolumn{5}{r}{{Continued on next page...}} \\
\endfoot

\bottomrule
\endlastfoot


拆字 & Character Decomposition & 把一个字或词拆成多个部件、偏旁或近似结构表达，使模型难以直接匹配原词。 & Splitting a Chinese character or word into components, radicals, or visually related parts to obscure the original expression. & “禾急贝兼” $\rightarrow$ “稳赚” \\
\addlinespace

同音 & Homophonic Substitution & 用读音相同或相近的字、词、拼音或谐音表达替换原始风险词。 & Replacing the original expression with characters, words, or phonetic forms that have the same or similar pronunciation. & “薇信” $\rightarrow$ “微信”; “稳砖” $\rightarrow$ “稳赚” \\
\addlinespace

形近 & Glyph Substitution & 用字形相近、视觉上容易混淆的字符替换原字符。 & Replacing characters with visually similar characters to preserve human readability while disrupting model recognition. & “很行” $\rightarrow$ “银行”; “己冻结” $\rightarrow$ “已冻结” \\
\addlinespace

繁简 & Traditional-Simplified Mixing & 使用繁体、异体字、简繁混用或其他字符变体替换规范简体表达。 & Mixing simplified, traditional, variant, or non-standard Chinese characters to disguise the canonical form. & “{\CJKfamily{bsmi}賺錢}” $\rightarrow$ “赚钱”; “{\CJKfamily{bsmi}發財}” $\rightarrow$ “发财” \\
\addlinespace

数混 & Alphanumeric \& Symbol Mixing & 使用数字、字母、符号与汉字混合表达, 或用形似字符替代原字符。 & Mixing digits, letters, symbols, and Chinese characters, or replacing characters with visually similar alphanumeric symbols. & “V信” $\rightarrow$ “微信”; “1oo\%” $\rightarrow$ “100\%”; “O” $\rightarrow$ “0” \\
\addlinespace

符号 & Symbol Insertion & 在词语内部或关键片段之间插入无意义符号, 破坏连续匹配。 & Inserting irrelevant symbols within words or key phrases to break lexical matching while keeping the text understandable. & “加-微-信” $\rightarrow$ “加微信”; “赚￥钱” $\rightarrow$ “赚钱” \\
\addlinespace

空格 & Whitespace Insertion & 在词语内部或关键片段之间插入空格、制表符或异常间隔。 & Inserting spaces, tabs, or abnormal gaps within words or key phrases to disrupt tokenization and matching. & “验\ 证\ 码” $\rightarrow$ “验证码”; “加\ 微\ 信” $\rightarrow$ “加微信” \\
\addlinespace

URL拆分 & URL Splitting & 对网址、域名、短链或联系方式进行空格、符号、换行等拆分。 & Splitting URLs, domains, short links, or contact strings with spaces, symbols, or line breaks to avoid URL-based detection. & “www . abc . com” $\rightarrow$
“www.abc.com”；“t\ .\ cn\ /
xx” $\rightarrow$ “t.cn/xx” \\
\addlinespace

前后缀 & Noise Prefix/Suffix & 在正文前后插入无意义字符、乱码、表情、符号或干扰短语。 & Adding meaningless characters, random symbols, emojis, or distracting phrases before or after the main message. & “@@@稳赚项目\#\#\#” $\rightarrow$ “稳赚项目” \\
\addlinespace

缩写 & Abbreviation & 使用拼音首字母、英文缩写、简称、符号化表达或平台黑称表示原词。 & Using initials, pinyin abbreviations, shorthand, symbolic forms, or aliases to refer to the original expression. & “+V” $\rightarrow$ “加微信”; “VX” $\rightarrow$ “微信” \\
\addlinespace

黑话 & Coded Jargon & 使用行业黑话、隐晦说法、暗语或上下文依赖表达来隐藏真实意图。 & Using domain-specific slang, euphemisms, coded expressions, or context-dependent phrases to conceal the real intent. & “上车” $\rightarrow$ “参与项目/入
局” \\
\addlinespace

多重 & Compound Obfuscation & 同一片段同时包含两种或多种伪装方式,需要联合判断才能还原。 & Applying multiple obfuscation strategies to the same span, requiring combined interpretation for restoration. & “加-薇-信V” $\rightarrow$ “加微信” \\
\addlinespace

未知 & Unknown Obfuscation & 存在明显伪装痕迹,但无法可靠归入以上具体类型,或还原结果不确定。 & Obfuscation is present, but its type cannot be reliably assigned to the predefined categories, or the restoration is uncertain. & 火星文/异常混写 $\rightarrow$ 还原结果不确定\\

\end{xltabular}
\end{CJK*}
\twocolumn 


\end{document}